\documentclass[%
 reprint,
superscriptaddress,
 amsmath,amssymb,
 aps,
]{revtex4-2}
\usepackage[colorlinks,citecolor=blue,linkcolor=blue,urlcolor=blue,bookmarks=false,hypertexnames=true
]{hyperref}
\usepackage{float}
\usepackage{tensor}
\usepackage{graphicx}
\usepackage{dcolumn}
\usepackage{bm}

\newenvironment{equations}{\equation\aligned}{\endaligned\endequation}  
\newcommand{\beq}{\begin{equations}}
\newcommand{\eeq}{\end{equations}}
\newcommand{\boe}{\begin{outline}[enumerate]}
\newcommand{\eo}{\end{outline}}
\def\l{\left(}
\def\r{\right)}

\def\la{\mathcal{L}}

\def\={&=}
\def\p{\partial}

\def\-{\item[-]}

\def\f21{\tensor[_2]{F}{_1}}
\def\m0{m_{0}}
\def\la0{\lambda_{0}}
\begin{document}

\preprint{APS/123-QED}

\title{On Aspects of Spontaneous Symmetry Breaking in Rindler and Anti-de Sitter Spacetimes for the $O(N)$ Linear Sigma Model}
\author{Pallab Basu}
\email{pallab.basu@wits.ac.za}
\affiliation{Mandelstam Institute for Theoretical Physics, University of Witwatersrand, Johannesburg, South Africa}
\author{S R Haridev}
\email{p20180460@hyderabad.bits-pilani.ac.in}
\author{Prasant Samantray}%
\email{prasant.samantray@hyderabad.bits-pilani.ac.in}
 \affiliation{BITS- Pilani Hyderabad Campus, Jawahar Nagar, Shamirpet Mandal, Secunderabad, 500078, India}


\begin{abstract}
We investigate aspects of spontaneous breakdown of symmetry for $O(N)$ symmetric linear sigma model in the background of Rindler and Anti-de Sitter spacetimes respectively. In the large $N$ limit, by computing the one-loop effective action, we report that in three dimensional Rindler space, there is a phase transition from the disordered phase to an ordered phase past a certain critical Rindler acceleration parameter `$a$'. Connections with finite temperature field theory results are established, thereby further reinforcing the idea that Rindler space can indeed be a proxy for Minkowski spacetime with finite temperature. We extend our calculations to Anti-de Sitter space in various dimensions and observe that symmetry is broken in three dimensions, but not in four dimensions. We discuss the implications of our results.
\end{abstract}

\keywords{$O(N)$ Linear sigma model, Spontaneous symmetry breaking, Accelerated observer, Anti-de Sitter space}
                              
\maketitle

\section{Introduction}
 The study of Quantum field theory (QFT) in curved spacetimes has a long and illustrative history of nearly half a century, including path-breaking works like Hawking's derivation of black hole radiation \cite{Hawking:1971vc}. In these works, the gravitational back-reaction of quantum fields is often neglected, and gravity and the curved spacetime serve only as a classical arena where the quantum field resides. We will call this approximation scheme: a probe limit. However, even within the approximation scheme of a probe limit, important physics like black hole radiation is demonstrated. Here in this work, following the same general theme of QFT in a curved spacetime, we  would take up certain questions, namely in the context of spontaneous symmetry breaking and study these questions \cite{PhysRevD.7.1888,parker_toms_2009,Buchbinder:1992rb}.  Spontaneous symmetry breaking is not only the central pillar in our state of the art understanding of standard model \cite{Higgs:1964ia}, but the same physical mechanism serves as a bedrock for superconductivity and other related phenomena in condensed matter physics \cite{PhysRev.130.439}. Also, the breakdown of spontaneous symmetry and associated phase transitions play important roles in cosmology \cite{ADLinde_1979,Padmanabhan_1983,ALLEN1983228}. Details of such calculations in general curved spacetimes can be found in  \cite{ILBuchbinder_1985,Markkanen2018,PMoniz_1990,PhysRevD.48.2813,ELIZALDE1994199}.
 
 The precise model we would concentrate on is a $O(N)$ symmetric linear sigma model with $N$ (usally $N \gg 1$) scalar fields. The spontaneous symmetry breaking in this model was studied analytically in flat spacetime \cite{colemanbook}. It is to be mentioned that large $N$ $O(N)$ sigma models are one of the few examples where dynamical symmetry breaking could be studied non-perturbatively in the $\phi^4$ coupling. 
 Actually a large $N$ saddle point calculation shows a breaking of $O(N)$ symmetry in three dimensional linear sigma models. 
 Here, we would like to study how the physics of large-$N$ sigma models are affected by curved  backgrounds and spacetimes with an event horizon.  
  
  To that end, we focus on Anti-de Sitter (AdS space) and Rindler space for concreteness. AdS spaces have been used heavily in the context of gauge-gravity correspondence, or holography \cite{Maldacena_1999}. Even in the context of holography, QFT phenomena like superconductivity have already been modeled/discussed in a probe limit \cite{https://doi.org/10.48550/arxiv.1612.07324}. Another spacetime we would look at is the Rindler space. Despite not being curved, Rindler space is interesting in its own right as it possesses an event horizon and, as an extension - a temperature as well (Fulling–Davies–Unruh effect \cite{Davies_1975,PhysRevD.14.870}). Additionally, it is to be noted Rindler space and AdS appear as a near-horizon geometry of non-extremal and extremal black holes, respectively.  

The plan of the paper is as follows: In section \ref{sec2}, we outline the $O(N)$ symmetric sigma model in large $N$ limit for an arbitrary spacetime. In sections \ref{sec3} and \ref{sec4}, we calculate the one loop effective potential for Rindler and in Anti-de Sitter spacetimes in various dimensions and consider scenarios for spontaneous symmetry breaking/restoration and phase transitions. Finally, we summarize our results in section \ref{sec5} and discuss their implications.  
\section{O(N) symmetric linear sigma model in the large $N$ limit}\label{sec2}
The $O(N)$ symmetric scalar field with $\lambda\phi^{4}$ type interaction is described by the Euclidean action \citep{colemanbook,coleman,4dlsm,4dlsm2,PhysRevD.50.5137}
\beq
S =\int dv_{x}\;  \l\frac{1}{2}g^{\mu\nu}\p_{\mu}\phi\p_{\nu}\phi +\frac{1}{2}m_{0}^{2}\phi^{2}+\frac{\lambda_{0}}{8N}\l\phi^{2}\r^{2}\r,
\eeq 
where $\phi$ is an $N$-component field, $\phi^{2}=\phi.\phi$ and $dv_{x} = d^{d}x\sqrt{g}$ is the $d$ dimensional invariant measure.  One can introduce an auxiliary field to the theory which does not affect the dynamics of the field as \citep{coleman,colemanbook}
\beq\label{scl}
S= \int dv_{x}\Bigg(\frac{1}{2}\phi\l-\nabla^{2}+\sigma\r\phi  -\frac{N}{2\lambda_{0}}\sigma^{2}+\frac{N m_{0}^{2}}{\lambda_{0}}\sigma\Bigg),
\eeq
where $\nabla^{2}$ is the $d$ dimensional Laplacian. Then the partition function for the theory is
\beq\label{z}
Z \= \int D\phi D\sigma~e^{-S[\phi,\sigma]} .
\eeq 
The path integral over $\phi$ in Eq. \ref{z} can be evaluated using the standard Gaussian integral, which gives
\beq
Z \= \int D\sigma\; e^{-\frac{N}{2}\log\l\det\l-\bar{\nabla}^{2}+\bar{\sigma}\r\r -\int dv_{x}\l-\frac{N}{2\lambda_{0}}\sigma^{2}+\frac{N m_{0}^{2}}{\lambda_{0}}\sigma\r}\\
\= \int D\sigma\; e^{-S_{eff}},
\eeq
where  $\bar{O}= O/\mu^{2}$ for any symbol `$O$' and with $\mu$ as some arbitrary constant with the dimension of mass and $S_{eff}$ is the effective action and is given as
\beq
S_{eff}=& \frac{N}{2}\log\l\det\l-\bar{\nabla}^{2}+\bar{\sigma}\r\r \\
&+\int dv_{x}\l-\frac{N}{2\lambda_{0}}\sigma^{2}+\frac{N m_{0}^{2}}{\lambda_{0}}\sigma\r\\
=& \frac{N}{2}\int dv_{x}\; \log\l -\bar{\nabla}_{x}^{2}+\bar{\sigma}\r \\
&+\int dv_{x}\l-\frac{N}{2\lambda_{0}}\sigma^{2}+\frac{Nm_{0}^{2}}{\lambda_{0}}\sigma\r.
\eeq
If the theory consists of a large number of scalar fields ($N\rightarrow\infty$), the dominant contribution to $Z$ comes from the saddle point of  $S_{eff}$. One can evaluate the effective action and the corresponding effective potential ($V_{eff}$) around the saddle point as
\beq\label{veff}
V_{eff}\= \frac{N}{2}\log\l -\bar{\nabla}_{x}^{2}+\bar{\sigma}\r -\frac{N}{2\lambda_{0}}\sigma^{2}+\frac{N m_{0}^{2}}{\lambda_{0}}\sigma.\\
\eeq
Note that quantum corrections does not involve $\m0$ or $\la0$ in it. We can rewrite the logarithm of an operator using the coincident limit of the corresponding Euclidean Green's function as \citep{PhysRevD.12.965,PhysRevD.105.105003,ooguri,BURGESS1985137,birrell_davies_1982}
\beq\label{v1}
V_{1}\=  \frac{N}{2}\log\l -\bar{\nabla}_{x}^{2}+\bar{\sigma}\r\\
\= \frac{N}{2}\int_{0}^{\sigma}dm^{2}\; \lim_{u\rightarrow 0}G\l u;m^2\r.
\eeq
 Here $G\l u,m^2\r$ is the Euclidean propagator of a massive scalar field of mass $m$ in $d$ dimensional space and $u$ is the invariant distance.  In the coincident limit ($u\rightarrow 0$) propagator diverges. For a renormalizable theory, looking at the classical action (Eq. \ref{scl}) we can write the renormalization conditions as
\begin{subequations}\label{12}
\begin{align}
\frac{d V_{eff}}{d\sigma}\Big|_{\sigma\rightarrow \mu^{2}} &= \frac{N m^{2}}{\lambda} \label{1},\\
\frac{d^{2}V_{eff}}{d\sigma^{2}}\Big|_{\sigma\rightarrow \mu^{2}} &= -\frac{N}{\lambda},\label{2}
\end{align}
\end{subequations}
where $\lambda$ and $m$ are the renormalized parameters. Here $\mu$ is an arbitrary constant with dimensions of energy and in possible cases we take $\mu =0$.  After renormalization, the total potential ($V$) is 
\beq\label{v}
V = \frac{1}{2}\sigma\phi^{2}+V_{eff}^{ren},
\eeq
where $V_{eff}^{ren}$ is the renormalized effective potential. The extreme of the total potential is determined by stationary points of $V$, $i.e.$,
\begin{subequations}\label{con}
\begin{align}
\frac{\p V}{\p \sigma} \= 0\label{con1},\\
\frac{\p V}{\p\phi}\=0\label{con2}.
\end{align}
\end{subequations}
We can use Eq.  \ref{con1} for writing $\sigma$ as a function of $\phi$ and can use the result to write $V$ as  a function of $\phi$ alone.  Then the condition for extrema of the potential is
\beq\label{vdphi}
\frac{dV}{d\phi} \= \frac{\p V}{\p \phi} + \frac{\p V}{\p\sigma}\frac{\p\sigma}{\p\phi}\\
\= \frac{\p V}{\p\phi}= 0,
\eeq
where in the second line of Eq. \ref{vdphi} we use Eq. \ref{con1}.  Using Eq. \ref{vdphi} in Eq. \ref{v} gives
\beq\label{extrema}
\frac{dV}{d\phi}= \sigma\phi .
\eeq
Now the minima of the potential can occur at $\phi=0$ or $\sigma=0$. The field configuration corresponds to  $\phi=0$ minima is $O(N)$ symmetric. But the field configuration corresponds to $\sigma =0$ is not $O(N)$ symmetric. So, if the global minima of the potential are at $\sigma=0$, then one says the theory exhibits spontaneous symmetry breaking \cite{PhysRev.127.965,PhysRev.122.345,zee}.

The formalism outlined here is generic. The $O(N)$ symmetric linear sigma model at large $N$ limit in Minkowski spacetime is well studied in the literature \cite{coleman,4dlsm,4dlsm2}. In flat spacetime, the $O(N)$ symmetric linear sigma model exhibits spontaneous symmetry breaking in three dimensions \cite{coleman}, and there is no spontaneous symmetry breaking in two, and four dimensions \cite{coleman,4dlsm,4dlsm2}.
\section{Rindler space}\label{sec3}
We now focus on studying spontaneous symmetry breaking of $O(N)$ symmetric linear sigma model in the large $N$ limit in Rindler space. Since Rindler space is the spacetime, as perceived by an uniformly accelerated observer in Minkowski spacetime, the idea is to check if observer dependence plays a role in the effects of spontaneous symmetry breaking. For the case of $N=1$, $\lambda\phi^{4}$ theory, the spontaneously broken $\mathrm{Z}_2$ symmetry is restored for a uniformly accelerated observer at a certain critical acceleration \citep{Padmanabhan_1983,Castorina2012-nb,rindlerd}. To check whether similar behavior persists in the large $N$ limit for the linear sigma model, we start by considering the theory in Rindler coordinates \cite{rindler,fulling}. The Rindler metric for a $d+2$ dimensional space with Euclidean signature is given as
\beq\label{rindlermetric}
ds^{2} = \xi^{2}d\tau^{2}+d\xi^{2} + (dx^{1})^{2}+...+(dx^{d})^{2}.
\eeq
Note that in this choice of coordinates,  $\tau$ is dimensionless.  As we are interested in the qualitative behavior of the effective potential, all one has to do is to read off the results from \cite{rindlerd} with $M^{2}$ replaced with $\sigma$ as
\beq\label{approxrindlervd}
V_{eff}^{ren}\= \frac{N}{2\lambda}\sigma^{2}+\frac{N m^{2}}{\lambda}\sigma -\frac{2N \sigma^{\frac{d+1}{4}}}{\l 4\pi\r^{\frac{d+2}{2}}\pi^{\frac{3}{2}}\xi^{\frac{d+3}{2}}}K_{\frac{d+1}{2}}\l 2\alpha\r,
\eeq
where $m$ and $\lambda$ are the renormalized parameters and $\alpha = \xi\sqrt{\sigma}$ which is a dimensionless parameter. With the effective potential in hand, one can study the symmetry behavior of the vacuum configurations in different dimensions. 
\subsection{Four dimensions}
In four dimensions, the near-horizon limit of the effective potential is
\beq\label{4dnearhorizon}
\lim_{\alpha\rightarrow 0}V_{eff}^{ren}\Big|_{d=2} =  -\frac{N}{2\lambda}\sigma^{2}+\frac{N m^{2}}{\lambda}\sigma  + \frac{N a^{2}}{16\pi^{3}}\sigma,
\eeq
where we choose the trajectory $\xi=1/a$. Also, one can use Eq. \ref{approxrindlervd} with $d=2$ in renormalization condition (Eq. \ref{1}) to see the effective mass as
\beq
m_{eff}^{2}= m^{2}+\frac{a^{2}}{16\pi^{3}},
\eeq
where there is a correction to the mass squared due to acceleration. However, the effective potential for the linear sigma model in large $N$ limit is double valued in four dimensional Minkowski space and the global minima of the potential is symmetric under $O(N)$ transformation for all values of $m$ and $\lambda$ \cite{4dlsm,4dlsm2}. So, there is no spontaneous symmetry breaking for linear sigma model in large $N$ limit in standard flat space. Therefore, one can not comment on the frame dependence of spontaneous symmetry breaking by naively looking at Eq. \ref{4dnearhorizon} in four dimensions. 

\subsection{Three dimensions}
It is known in the literature \cite{coleman} that in three dimensions with $m^{2}<0$,  the vacuum configuration of the field is not symmetric under $O(N)$ transformations. This implies that $O(N)$ symmetry is spontaneously broken. Similar to the $\lambda\phi^{4}$ theory \cite{rindlerd}, one can therefore expect the restoration of $O(N)$ symmetry in accelerated frames. We can study symmetry restoration by considering the near horizon limit of the effective potential. In three dimensions ($d=1$ in Eq. \ref{approxrindlervd}), the near-horizon limit of the effective potential is (leading order in $a$)
\beq\label{3dnearhorizon}
\lim_{\alpha\rightarrow 0}V_{eff}^{ren}\Big|_{d=1} =  -\frac{N}{2\lambda}\sigma^{2}+\frac{N m^{2}}{\lambda}\sigma -\frac{N\sigma a}{8\pi^{3}}\log\l\frac{\sigma}{a^{2}}\r,
\eeq
where we choose the trajectory $\xi=1/a$. Now consider a theory with spontaneous symmetry breaking in standard flat space ($m^{2}<0$). In accelerated frames (from Eq. \ref{3dnearhorizon}), the effective mass (from Eq. \ref{1}) turns positive after some critical acceleration or the symmetry violation disappear for some high acceleration. Now one can try to calculate the effective mass and the corresponding critical acceleration using Eq. \ref{approxrindlervd} with $d=1$ in Eq. \ref{1}. But the critical acceleration will depend on some arbitrary scale $\mu$ (despite the fact that the normalized effective potential is independent of any such scale). So, like in finite temperature field theory \cite{Fujimoto1987-fb}, Eq. \ref{3dnearhorizon} is not very useful for predicting the critical acceleration. However, the structure of Eq. \ref{3dnearhorizon} guarantees that symmetry is indeed restored for at least a certain class of Rindler observers. Additionally, our techniques are general and can be utilized to comment on the aspects of symmetry breaking in Rindler space for arbitrary spacetimes.

\section{Anti-de Sitter space}\label{sec4}
We now consider a different physical scenario. We inquire whether the curvature of spacetime plays a role in the effects of phase transition or symmetry breaking/restoration for the linear sigma model. To that end we consider the $O(N)$ linear sigma model in the background of Anti-de Sitter spacetime focusing on three and four dimensions. We consider Euclidean $AdS_{d+1}$ with the metric
\beq\label{adsmetric}
ds^{2} = \frac{L^{2}}{z^{2}}\l dz^{2}+ \sum_{i=1}^{d}dx_{i}^{2}\r,
\eeq
where $L$ is the $AdS$ scale. The bulk to bulk scalar Green's function for $AdS_{d+1}$ is well discussed in the literature \cite{maldacena,ooguri,BURGESS1985137}, and we borrow the results from \citep{maldacena} as
\beq\label{prop}
G(W)\=\frac{\alpha_{0}}{L^{d-1}}W^{\Delta}\f21\l\Delta,\Delta+\frac{1-d}{2},2\Delta-d+1;-4W \r,
\eeq
where
\beq\label{prop1}
W \= \frac{1}{2}\frac{1}{\cosh(\frac{u}{L})-1},\\
\Delta \= \frac{d}{2}+\frac{1}{2}\sqrt{d^{2}+4m^{2}L^{2}},\\
\alpha_{0}\= \frac{\Gamma(\Delta)}{2\pi^{d/2}\Gamma\l \Delta-\frac{d}{2}+1 \r},
\eeq
and $\f21$ is the usual Hypergeometric function. Note that in the coincident limit (i.e., $u\rightarrow 0$), $W$ diverges, and Green's function is singular as expected. We now specialize to three and four dimensions for our analysis.
\subsection{Four Dimensions}
One can expand $G(W)$ around $u=0$, for the case of $d=3$ this gives
\beq\label{series}
\lim_{u\rightarrow 0}G(W) \= \frac{1}{12\pi^{2}}\l \frac{3}{u^{2}} -\frac{16+3\Delta(\Delta-5)}{4L^{2}}+\frac{3}{4L^{2}}\times\right.\\
&\left.\l\Delta-2\r\l\Delta-1\r\l 2H_{\Delta-3}+ 2\log\l\frac{u}{2L}\r\r\r,
\eeq
where $H_{z}$ is the harmonic number. Using Eq. \ref{series} we can calculate $V_{1}$ (see Eq. \ref{v1}) as
\beq\label{v1ads2}
V_{1}\= \frac{N}{8L^{2}\pi^{2}}\l\sigma\l \gamma + \frac{L^{2}}{u^{2}} -\frac{5}{12} +\frac{\beta}{3}+\log\l\frac{u}{2L}\r\right.\right.\\
&\left.\left.+\frac{1}{2}\log\l\Gamma(\beta)\r+ \beta\log\l\Gamma(\beta)\r - 3\psi\l -2,\beta\r \r\right.\\
&\left.+\frac{L^{2}\sigma^{2}}{4}\l \log\l\frac{u}{2L}\r +\gamma -\frac{1}{2}\r -\frac{13}{2L^{2}}\psi\l -2,\beta\r\right.\\
&\left.+ \frac{\beta}{L^{2}}\l \frac{3}{4} + 2\log\l\Gamma(\beta)\r + 6\psi\l -3,\beta\r\r\right.\\
&\left.+\frac{1}{L^{2}}\l \log\l\Gamma(\beta)\r - 6\psi\l -4,\beta\r + 3\psi\l -3,\beta\r \r\r,
\eeq
where $\gamma$ is the Euler's constant, $\psi$ is the polygamma function and for notational simplicity we choose 
\beq
\beta = -\frac{1}{2}+\frac{1}{2}\sqrt{9 + 4 L^{2}\sigma}.
\eeq
Eq. \ref{v1ads2} regularizes the effective potential and the divergences in the effective potential is in linear and quadratic powers of $\sigma$ \citep{BURGESS1985137,ooguri} allowing us to renormalize the theory by redefining $m_{0}$ and $\lambda_{0}$. Using the renormalization condition (Eq. \ref{1}), we can write
\beq\label{4drencon1}
\frac{1}{48L^{2}\pi^{2}}+\frac{1}{8\pi^{2}u^{2}}+\frac{m_{0}^{2}}{\lambda_{0}}+\frac{1}{8L^{2}\pi^{2}}\log\l\frac{u}{2L}\r=\frac{m^{2}}{\lambda},
\eeq
and Eq. \ref{2} gives
\beq\label{4drencon2}
\frac{1}{144}-\frac{1}{96\pi^{2}}-\frac{1}{\lambda_{0}}+\frac{1}{16\pi^{2}}\log\l\frac{u}{2L}\r =-\frac{1}{\lambda}.
\eeq
Using Eq. \ref{4drencon1} and Eq. \ref{4drencon2} one can write the renormalized effective potential as
\beq\label{veffren}
V_{eff}^{ren}\= \frac{Nm^{2}\sigma}{\lambda}-\frac{N\sigma^{2}}{2\lambda}+\frac{N}{8L^{2}\pi^{2}}\l\frac{\sigma^{2}L^{2}}{4}\l \gamma -\frac{1}{3}-\frac{\pi^{2}}{9}\r\right.\\
&\left.+\frac{\sigma}{2}\l 2\gamma + \frac{2\beta}{3}+\l 2\beta +1\r\log\l\Gamma(\beta)\r-6\psi\l -2,\beta\r\right.\right.\\
&\left.\left.-\frac{7}{6}\r+ \frac{\beta}{L^{2}}\l \frac{3}{4}+2\log\l\Gamma(\beta)\r + 6\psi\l -3,\beta\r\r\right.\\
&\left.+\frac{1}{L^{2}}\l \log\l\Gamma(\beta)\r -6\psi\l -4,\beta\r + 3\psi\l -3,\beta\r\right.\right.\\
&\left.\left.-\frac{13}{2}\psi\l -2,\beta\r\r\r.
\eeq
Now using Eq. \ref{v} one can calculate the total potential. As a check, in the limit $L\rightarrow \infty$, $V$ matches with the flat space potential as \citep{coleman,4dlsm,4dlsm2}
\beq
\lim_{L\rightarrow\infty}V = \frac{1}{2}\sigma\phi^{2}+\frac{m^{2}N\sigma}{\lambda}-\frac{N\sigma^{2}}{2\lambda}+\frac{N\sigma^{2}}{64\pi^{2}}\log\l\frac{\sigma}{\mu^{2}}\r,
\eeq
where we choose $\mu=1/L$ as the arbitrary energy scale.
\subsubsection{Ground state}
In order to understand spontaneous symmetry breaking, we begin by determining the stationary points of the potential $V$. Substituting Eq. \ref{veffren} in Eq. \ref{con1} gives
\beq
\phi^{2}(\sigma)\= \frac{N}{72L^{2}\pi^{2}\lambda}\Bigg( -144\pi^{2}L^{2}m^{2}+3\lambda L^{2}\sigma+\lambda\pi^{2}L^{2}\sigma\\
&+144 L^{2} \pi^{2}\sigma-9\lambda\l\beta +1+L^{2}\sigma\r H_{\beta-1}\Bigg).
\eeq
\begin{figure}[H]
\centering
\includegraphics[scale=0.5]{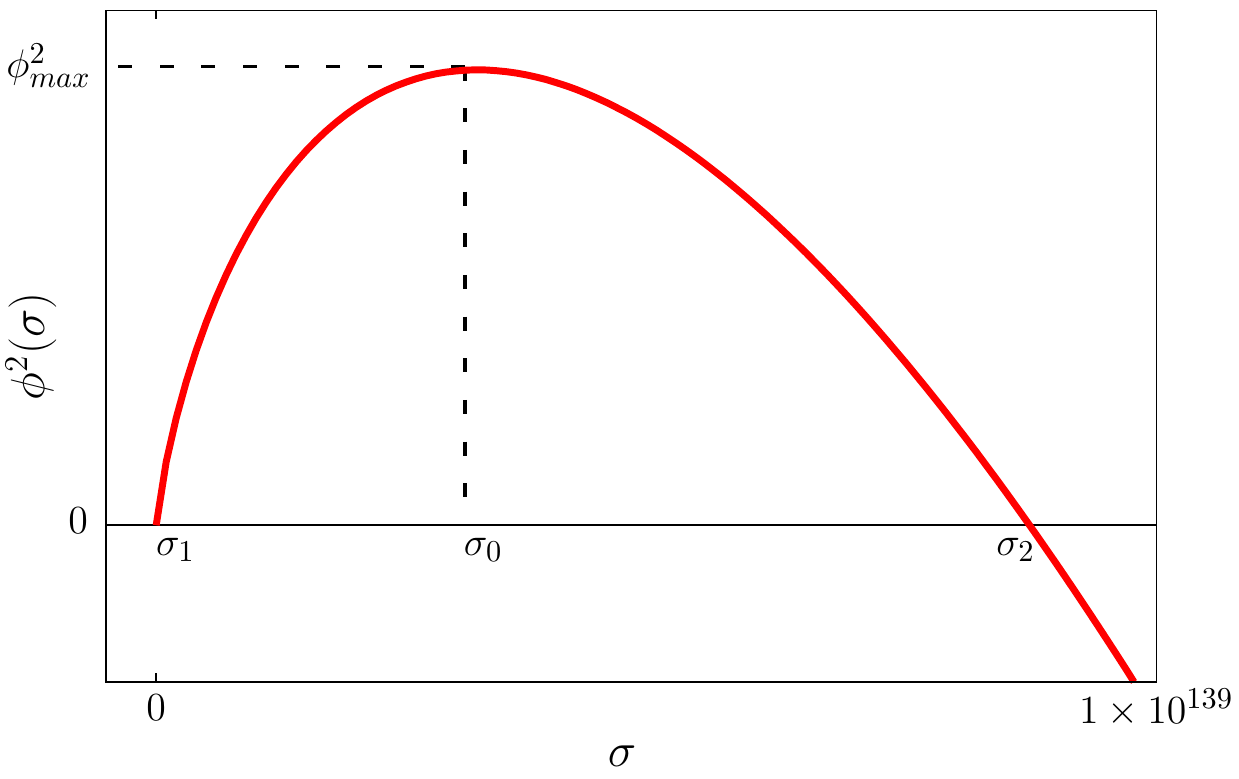}
\caption{A plot of $\phi^{2}$ versus $\sigma$ for $m^{2}>0$.}
\label{fphi2}
\end{figure}
The variation of $\phi^{2}$ as a function of $\sigma$ is shown in Fig. \ref{fphi2}.  From Fig. \ref{fphi2} we can see that $\phi^{2}$ attains a maximum value $\phi_{max}^{2}$ at $\sigma =\sigma_{0}$ and monotonically decreases after that. This is similar to the case in four dimensional flat space and it indicates that $V(\phi^{2})$ is a double valued function of $\phi^{2}$ for $\phi^{2}<\phi^{2}_{max}$ \cite{4dlsm,4dlsm2}.

The field configuration with minimum potential determines the ground state of the theory. One can find the ground state by comparing the values of the potential at possible minima. Consider the minima at $\sigma = 0$
\beq
V(0) = -\frac{0.003N}{L^{4}}.
\eeq
But at this point
\beq
\phi^{2}(0) = -\frac{2m^{2}N}{\lambda}.
\eeq
 For $m^{2}>0$ the field becomes complex, therefore $\sigma=0$ requires $m^{2}\leq 0$. From Fig. \ref{fphi2}, we can see that $\phi=0$ for both $\sigma_{1}$ and $\sigma_{2}$.  For $m^{2}<0$, and $\phi^{2}(0)>0\implies \sigma_{1}<0$ which is not in the possible range of $\sigma\in [0,\infty)$.  So the only possible other minima of the potential is at $\sigma_{2}$. From Eq. \ref{v}
\beq
V(0) \= V_{eff}^{ren}(\sigma).
\eeq
For small values of $\sigma$
\beq\label{vsigma0}
\lim_{\sigma\rightarrow 0}V_{eff}^{ren}(\sigma) = V(0) + \frac{Nm^{2}\sigma}{\lambda} + O(\sigma^{2}).
\eeq
So for $m^{2}<0$, $V_{eff}$ decreases from $V(0)$ to $V(\sigma_{2})$. Then
\beq
V(\sigma_{2})< V(0).
\eeq
So,  for $m^{2}<0$ the global minima of the potential is at $\phi=0$. Therefore, the global minima of the potential is a symmetric one and there is no spontaneous symmetry breaking.

Now, if $m^{2}>0$ the minimum of the potential can only occur at $\phi=0$.  Subsequently, there is no spontaneous symmetry breaking in the system in this case.  Now the minimum of the potential can be at $\sigma_{1}$ or $\sigma_{2}$.  For $m^{2}> 0$, as a function $V_{ren}^{eff}$ starts to increase from $\sigma=0$ (from Eq. \ref{vsigma0}) and reaches its first extremum at $\sigma_{1}$ and then decreases till $\sigma_{2}$. So
\beq
V(\sigma_{2})<V(\sigma_{1}),
\eeq
which makes $\sigma_{2}$ the global minima.  Now for $m^{2}=0$ we can repeat the same arguments for $m^{2}<0$ and conclude that $\phi(\sigma_{2})$ is the global minimum  configuration of the field.

In conclusion, in four dimensional $AdS$ there is no spontaneous symmetry breaking for any values of $m^{2}$ with $\lambda>0$.
\subsection{Three Dimensions}
The computation for three dimensional case is similar to that of four dimensions.  In three dimension,  we can expand the propagator near $u=0$ as (from Eq. \ref{prop})
\beq
\lim_{u\rightarrow 0}G(W) = \frac{1}{4\pi L}+\frac{1}{4\pi u}-\frac{\Delta}{4\pi L} + O(u),
\eeq
where $\Delta$ is defined in Eq. \ref{prop1}. Substituting this in Eq. \ref{v1}  one can calculate $V_{1}$ as
\beq
V_{1}= \frac{N\sigma}{8\pi u} -\frac{N}{12L\pi}\l\sigma +\frac{1}{L^{2}}\r\sqrt{1+L^{2}\sigma}.
\eeq
The divergence in $V_{1}$ is proportional to linear power in $\sigma$. So, one can renormalize it using Eq. \ref{1} as
\beq\label{3dcon1}
\frac{1}{8\pi}\l\frac{1}{u}+\frac{1}{L}\r + \frac{m_{0}^{2}}{\lambda_{0}} = \frac{m^{2}}{\lambda}.
\eeq
Using this condition (Eq. \ref{3dcon1}), the renormalized effective potential is
\beq
V_{eff}^{ren} \=\frac{N\sigma}{8\pi L}+\frac{Nm^{2}\sigma}{\lambda}-\frac{N\sigma^{2}}{2\lambda} \\
&-\frac{N}{12L\pi}\l\sigma +\frac{1}{L^{2}}\r\sqrt{ 1+L^{2}\sigma}.
\eeq
\subsubsection{Ground state}
The stationary points of the potential is determined by Eq. \ref{con}. Using Eq. \ref{con1} one can get
\beq
\phi^{2}(\sigma) = -\frac{N}{4\pi L}-\frac{2Nm^{2}}{\lambda}+\frac{2N\sigma}{\lambda}+\frac{N}{4\pi L}\sqrt{ 1+L^{2}\sigma}.
\eeq
Here
\beq
\frac{d\phi^{2}}{d\sigma}>0,
\eeq
which makes $\phi^{2}$ a monotonically increasing function of $\sigma$.  From Eq. \ref{extrema}, the stationary points of the potential $V$ occurs at $\phi=0$ or $\sigma=0$.  Consider the extremum at $\sigma=0$
\beq
V(0) =-\frac{N}{12\pi L^{2}},
\eeq
For which we have
\beq\label{3dphi2sigma0}
\phi^{2}(0) =-\frac{2Nm^{2}}{\lambda}.
\eeq
So value of $\sigma =0$ is possible only if $m^{2}\leq 0$. If we consider $m^{2}<0$ then from Eq. \ref{3dphi2sigma0} $\phi^{2}(0)>0$. A monotonically increasing function cannot reach zero starting from a positive value. Therefore, $\phi=0$ is not possible and the only possible extremum is at $\sigma=0$. As the global minima is an asymmetric one, $O(N)$ symmetry is spontaneously broken.  For $m^{2}>0$ and $m^{2}=0$, only possible ground state is at $\phi=0$ which is a symmetric one.

\section{Results and Discussion}\label{sec5}
We calculated the one loop effective potential for the $O(N)$ symmetric sigma model in Rindler space in arbitrary dimensions in the large $N$ limit (Eq. \ref{approxrindlervd}). Interestingly we observe that in three dimensions, the broken $O(N)$ symmetry will be restored in accelerated frames after a critical acceleration (Eq. \ref{3dnearhorizon}). This result is analogous to that of the finite temperature results \cite{Fujimoto1987-fb,EINHORN1993611}. We also observe that in four dimensions, the effective mass squared of the theory gets quantum corrected due to the existence of an acceleration parameter `$a$' having dimensions of mass. One can obtain similar conclusions for $N=1$ theory as shown in \cite{rindlerd}. The method for calculating the effective potential in Rindler spacetime is for arbitrary dimensions. So one can use our results for studying similar phenomenons in arbitrary dimensions.

These results are indicative of the fact that non-perturbative phenomena like spontaneous symmetry breakdown/restoration is possibly observer-dependent, and different Rindler observers report different critical values at which phase transition occurs. Similar conclusions are obtained for standard model phase transitions in \cite{PhysRevD.99.125018}. These non-trivial results can also be mapped to the presence of the event horizon in Rindler space which in itself is observer-dependent, and has a temperature associated with it. In this sense, a lot of field-theoretic results in Rindler space are in one-one correspondence with finite temperature physics \cite{Padmanabhan_1983}.

We performed similar calculations in the Anti-de Sitter space background, and qualitative results are similar to that obtained in flat space \cite{4dlsm,4dlsm2,coleman}. In four dimensions, there is no spontaneous symmetry breaking for any values of $m^{2}$. However, in three-dimensional AdS, we have spontaneous symmetry breaking for $m^{2}<0$. Also, one loop quantum effects make corrections proportional to $1/L$ and $1/L^{2}$ to the effective mass squared of the theory in three and four dimensions, respectively. The effect of curvature, therefore, in this context, was to essentially modify the effective mass squared of the scalar fields. An interesting question immediately arises. If one were to instead consider Rindler-AdS spacetime \cite{Parikh2018} which is both curved as well as possesses an event horizon - what would be the fate of the ground state of the linear $O(N)$ sigma model in such a spacetime? Especially interesting would be the critical temperature at which symmetry restoration might occur and the form of quantum mechanical corrections to the mass squared terms. We of course reserve these questions for future work.
\begin{acknowledgments}
Research of P.S is partially supported by CSIR grant 03(1350)/16/EMR-II Govt. of India, and also by the OPERA fellowship from BITS-Pilani, Hyderabad Campus. H. R is supported by CSIR grant 03(1350)/16/EMR-II Govt. of India.
\end{acknowledgments}
\bibliography{reference}

\end{document}